\documentclass[12pt,a4paper,reqno]{article}
\usepackage{a4}
\usepackage{pstricks,pst-node,pst-plot}
\usepackage{amsmath}
\usepackage{amssymb}
\usepackage{amsthm}

\newtheorem{lem}{Lemma}[section]

\newcommand{\tr}{\textrm{tr}}
\newcommand{\ra}{\rangle}
\newcommand{\la}{\langle}
\newcommand{\re}{\textrm{Re}}

\newcommand{\II}{\leavevmode\hbox{\small1\kern-3.8pt\normalsize1}}

\newcommand{\inbar}{\,\vrule height1.5ex width.4pt depth0pt}
\newcommand{\CC}{\relax{\hbox{$\inbar\kern-.3em{\rm C}$}}}

\newcommand{\ZZ}{\hbox{\sansf Z\kern-0.4em Z}}

\font\sansf=cmss12

\newpsobject{showgrid}{psgrid}{subgriddiv=1,griddots=5,gridlabels=6pt}

\begin{document}

\title{On maximal entanglement between two pairs 
in four-qubit pure states}
\author{\\\\S.~Brierley$^1$ and A.~Higuchi$^2$\\
\small \emph{Dept. of Mathematics, University of York, 
Heslington, York, YO10 5DD, U.K.}
\\\small $^1$email: sb572@york.ac.uk 
\\\small $^2$email: ah28@york.ac.uk}

\date{23 April, 2007}
\maketitle
\begin{abstract}
We show that the state with the highest known average two-particle 
von Neumann entanglement entropy proposed by Sudbery and one of the 
authors
gives a local maximum of this entropy. We also show that this is not the 
case for an alternative highly entangled state proposed by 
Brown et al.
\end{abstract}

		\section{Introduction}
The characterisation of multi-particle entanglement is a 
major open problem that is particularly significant for the study of
quantum computation and many-body physics~\cite{NC02}. One of the ways in which entanglement
can be understood is by way of reference to a ``maximally'' entangled state.
This target state can then be used, for example, to determine the largest rate at which it is
possible to distill pure maximally entangled states from a supply of
mixed states using only LOCC~\cite{PV07}. An obvious condition 
for maximal entanglement in the case of pure states is that 
all one-qubit reduced density matrices are maximally mixed. 
For two- and three-qubit systems
this leads to a unique state~\cite{SM96} (up to local unitary operations).
However this is not true for a system with more than three qubits. 
For example, the states 
\begin{align*}
|\phi_1 \ra 	&= \frac {1}{\sqrt{2}} ( |0000\ra +|1111\ra ), \\ 
|\phi_2 \ra 	&= 
\frac {1}{2} ( |0000\ra +|0111\ra +|1001\ra +|1110\ra ),
\end{align*}
have the property that all one-party 
reduced density matrices are maximally mixed;
yet these two states are not locally equivalent. One may then 
ask which states also have maximally mixed two-party reduced density
matrices.  However, 
Sudbery and one of the authors (AH) 
have shown that it is not possible for all two-qubit reduced density 
matrices of a pure four-qubit state
to be maximally mixed~\cite{HS00}.  Nevertheless, they found a state
which appears to maximise 
the average von Neumann entropy 
of two-qubit reduced density matrices, which will be denoted by $E_2$ in
this paper.
They showed that this state is a stationary point of the function
$E_2$, but it is not known whether this state indeed gives the maximum
of $E_2$.  In this paper we show that this state gives at least a
{\em local} maximum of $E_2$.

Given four qubits $A,B,C$ and $D$ 
the von Neumann entropy of the two-party reduced states is
\begin{equation}
E_{XY}=-{\rm tr}(\rho_{XY} \log_{2} \rho_{XY}),\label{EXY}
\end{equation}
where $\rho_{XY}=\tr _{ZW} | \psi \ra \la \psi |,$ with ${W,X,Y,Z}$ being a permutation
of the four systems. The average entropy of two-qubit reduced density
matrices is defined by
\begin{align*}
E_{2} 
& \equiv  \frac{1}{6} ( E_{AB}+E_{AC}+ E_{AD}+ E_{BC}+ E_{BD}+ E_{CD} ) \\
						& =\frac{1}{3} ( E_{AB}+E_{AC}+ E_{AD} ).
\end{align*}
This quantity can naturally be taken as a measure of the entanglement
between two pairs
in pure four-qubit states.  

There are other approaches to quantifying multi-partite entanglement that are applicable to a
general mixed state.  A measure for a general composite system has
been introduced by Yukalov~\cite{Yukalov} using the ratio of 
norms of an entangling operator
and of a disentangling operator in the relevant disentangled Hilbert space.
The entanglement
of a four-qubit system can also be studied using the entropy of the 
reduced three-particle
system and the strong subaddativity inequality~\cite{BA03}. 
Brown et al. have considered the
partial transpose with respect to all possible partitions 
of the state~\cite{Susan}.
These measures are more general than $E_{2}$
in so much as they are applicable to mixed states.
In this paper, we only consider
pure states and, therefore, the average entropy is a suitable measure of entanglement. 

The four-qubit state proposed in~\cite{HS00}, 
$$|M_{4} \ra = \frac{1}{\sqrt{6}} [ |0011\rangle+|1100\rangle+\omega(|0101\rangle+|1010\rangle)
+\overline{\omega}(|0110\rangle+|1001\rangle) ] ,$$
where $\omega = \exp(2i\pi/3)$ is a third root of unity, 
has the highest known average two-qubit bipartite entanglement~\cite{HS00, Susan}.
It is also an element of the orbit of SLOCC operations
that has maximal four-partite entanglement~\cite{Gabcd}. 
Together with its complex conjugate, it also provides
a basis for the space of singlets contained in a four-qubit 
Hilbert space~\cite{V4}.

The entropy $E_{XY}$ measures the entanglement between systems $XY$ and $WZ$, 
i.e. it measures entanglement between pairs.  The entropy of the one party reduced density
matrices measure the entanglement between individual systems and the rest of
the state. Another way in which entanglement could manifest
itself in a four qubit system is the entanglement between any two individual systems. For
example, the entanglement between $X$ and $Y$ is measured by regarding $\rho
_{XY}$ as a (mixed) state in its own right. There are various bipartite
entanglement measures for mixed states. One that is commonly used for
multipartite states is the concurrence, since for this measure the CKW
inequality gives an upper bound on the bipartite entanglement in terms of
one party entanglement \cite{CKW, OV06}. Namely, 
\begin{eqnarray}\label{concurrence}
C_{AB}^{2}+C_{AC}^{2}+C_{AD}^{2}\leq C_{A(BCD)}^{2},
\end{eqnarray}
where $C_{A(BCD)}$ denotes the concurrence across the partition $A:BCD$ i.e.
considering the qubits $BCD$ as one $8$ dimensional qudit. Since $\rho
_{ABCD}$ is pure, $C_{A(BCD)}=2\sqrt{\det (\rho _{A})}.$ 

It is interesting to note that $|M_{4}\rangle $ contains no bipartite
entanglement, that is, $\rho _{AY}$ is separable\footnote{$\rho _{AY}$ is a
mixed state, and so separability here means that it can be writen as the
convex sum of unentangled pure states.} for all $Y$. Moreover, the states $%
\rho _{AY}$ are on the boundary of separable states. This can be seen by
writing $\rho _{AY}$ in the form  
\begin{eqnarray}
\rho _{AY}=\frac{(1-\frac{1}{3})}{4}\mathbb{I}+\frac{1}{3}|\Phi _{-}\rangle
\langle \Phi _{-}|,
\nonumber
\end{eqnarray}
where $|\Phi _{-}\rangle =\frac{1}{\sqrt{2}}(|01\rangle -|10\rangle ),$ i.e.
as a Werner state with $x=\frac{1}{3}$ \cite{Peres,HHH96}. Therefore, the l.h.s. of eq. (\ref
{concurrence}) is zero while the r.h.s. is one. This can be thought of as saying
that $A$ shares none of it's entanglement with it's neighbours i.e. the
entanglement between $AY$ and $WZ$ that cannot be regarded as the
entanglements of $A$ with $W$ or $Z$ separately is maximised by $%
|M_{4}\rangle $.

The authors of~\cite{HS00} conjectured that $|M_{4} \ra$ gives a maximum of the
average two-qubit bipartite entanglement.  By considering the first variation, $\delta E_{XY}$, they
were able to show that $E_{2}$ is stationary at $|M_{4} \ra$. In this paper, we will
consider the second variation of the average entropy and demonstrate that
the state $|M_{4} \ra$ indeed gives
a local maximum. We will also consider another highly entangled state
proposed in~\cite{Susan} and show that this
state is in fact not a stationary point of our measure, thus 
illustrating that maximal entanglement is
dependent on the measure used. 

		\section{Varying the Entropy}

We consider variations of the state near $|M_4\ra$ with the varied states
characterised by several small parameters.  In general, suppose we vary
a four-qubit state $|\psi\rangle$ in second-order approximation as
$$
|\psi\rangle \rightarrow|\psi'\rangle
=|\psi\rangle+|\delta\psi\rangle+|\delta^{2}\psi\rangle,
$$
where $|\delta\psi\ra$ and $|\delta^2\psi\ra$ are of first order 
and of second order, respectively, 
in the small parameters characterising the
variations.  {}From the normalisation condition 
$\langle \psi'|\psi'\rangle = \langle \psi|\psi\rangle = 1$, we obtain
\begin{equation}\label{norm
1} 2\,\textrm{Re}\langle\delta\psi|\psi\rangle=0, \end{equation}
at first order and
\begin{equation}\label{norm 2} 2\,\textrm{Re}\langle \delta^{2}\psi|\psi\rangle+\langle
\delta\psi|\delta\psi \rangle=0, \end{equation}
at second order.

In order to find the first and second variations of the entropy, we need the following lemma.
\begin{lem}\label{var lem}
For any function $f(A)$ of a matrix $A$
that can be written as a power series, consider variations of $A$ with
small parameters, $A\mapsto A+\delta A + \delta^2 A$, to second order in
these parameters.  Then to second order the corresponding variation
${\rm tr}[f(A)] \mapsto {\rm tr}[f(A)] + \delta {\rm tr}[f(A)] 
+ \delta^2{\rm tr}[f(A)]$ is
given by 
\begin{align*}
\delta {\rm tr}[f(A)]     &={\rm tr}[\delta A\cdot f'(A )],\\
\delta^{2} {\rm tr}[f(A)] &={\rm tr}[\delta^{2}A\cdot f'(A )+\frac{1}{2}
\delta A\cdot \delta f'(A ) ].
\end{align*} 
\begin{proof}
Since $f(A)$ can be written as a power series, it is enough to
show these formulas for $f(A)=A^{n}$. 
We obtain first-order terms in the variation of $A^n$ 
by replacing one of the $A$'s by $\delta A$.  Thus,
$\delta \tr A^n = n\tr[\delta A\cdot A^{n-1}]$.  
This proves the first formula.
We get second-order terms by replacing one of $A$'s by
a $\delta^{2}A$, or by replacing two $A$'s by two $\delta A$'s. 
Thus,
\begin{eqnarray}
\delta^2 \tr A^n  &  = &  
n \tr[\delta^2 A\cdot A^{n-1}] + 
\frac{n}{2}
\sum_{k=0}^{n-2}\tr[\delta A\cdot A^k\cdot \delta A \cdot
A^{n-k-2}]\nonumber \\
& = & \tr\left[\delta^2 A\cdot nA^{n-1}\right]
+ \frac{1}{2}\tr\left[\delta A\cdot \delta (n A^{n-1})\right].
\nonumber 
\end{eqnarray}
(One can perhaps convince oneself of the need for the factor
$1/2$ in the second term by noting
that the number of ways to replace two $A$'s by two $\delta A$'s is 
$n(n-1)/2$.) 
This proves the second formula.
\end{proof}
\end{lem}
Letting $A=1-\rho_{XY}$ and $f(A) = (1-A)\log(1-A) =
\rho_{XY}\log\rho_{XY}$ in this Lemma, and using the normalisation 
conditions $\tr[\delta\rho_{XY}]=\tr[\delta^2\rho_{XY}] = 0$,
we obtain the first- and
second-order variations of $E_{XY}$ defined by (\ref{EXY}) as
\begin{eqnarray}
\delta E_{XY}  &=&-\frac{1}{\log2}\tr[\delta\rho_{XY}\log\rho_{XY}],
\label{firstorder}  \\
\delta^{2} E_{XY} &=&
-\frac{1}{\log2}\tr[\delta^{2}\rho_{XY} \log\rho_{XY}
+\frac{1}{2}\delta \rho_{XY} \,\delta \log\rho_{XY}], 
\label{secondorder}
\end{eqnarray}
where
\begin{align*}
\rho_{XY} &= \tr_{ZW}(|\psi\rangle\langle \psi|),\\
\delta\rho_{XY}&=\tr_{ZW}(|\delta\psi\rangle\langle
\psi|+|\psi\rangle\langle\delta \psi|), \\
\delta^{2}\rho_{XY}&=\tr_{ZW}(|\delta^{2}\psi\rangle\langle
\psi|+|\psi\rangle\langle\delta^{2}
\psi|+|\delta\psi\rangle\langle\delta \psi|). \end{align*}

	\section{An alternative highly entangled state}\label{alt}

In~\cite{Susan}, Brown et al. have considered multi-partite 
entanglement across all possible
partitions of a state and calculated 
the sum of all negative eigenvalues when the 
partial-transpose function is applied and sought to maximise 
this over all possible states.
In the four-qubit case, their numerical search found the state 
$$|\psi_{4}\ra = \frac{1}{2}
(|0000\ra+ |+011\ra+ |1101\ra+ |-110\ra),$$ where
$|+\ra = \frac{1}{\sqrt{2}}(|0\ra +|1\ra)$ and $|-\ra = \frac{1}{\sqrt{2}}(|0\ra -|1\ra)$. 
Here we consider whether the state $|\psi_{4}\ra$ gives 
a local maximum of $E_{2}$ as a measure of entanglement.
 
We start by noting that $\rho_{AC}$ is maximally mixed and that the eigenvalues of
$\rho_{AB}$
and $\rho_{AD}$ are both equal to 
$$\left \{
\frac{2+\sqrt{2}}{8},\frac{2+\sqrt{2}}{8},\frac{2-\sqrt{2}}{8},
\frac{2-\sqrt{2}}{8}  \right \}.$$ Comparing 
the average two-qubit entanglement of this state with that of $|M_{4}\ra$,
we see that the state $|\psi_4\rangle$
has lower two-party average von Neumann entanglement~\cite{Susan}. 
Namely, 
\begin{align*}
E_{2}( |\psi_{4}\ra)
&=\frac{5}{2} - \frac{1}{2\sqrt{2}} \log_{2}{(3+\sqrt{2})}\approx 1.7426, \\
E_2 ( |M_{4}\ra ) 		&=1+\frac{1}{2} \log_{2}{2}\approx 1.7925.
\end{align*}
We will now show that the state $|\psi_4\rangle$ 
in fact does not give a local maximum of $E_{2}$ by demonstrating that
it is not a stationary point.
The first-order variation of $E_{2}$ is given by
$$3\delta E_{2}=\delta E_{AB}+\delta E_{AC}+\delta E_{AD}.$$ 
Since $\rho_{AC}$ is maximally
mixed, we have $\delta E_{AC}=0$ by (\ref{firstorder}) and
the first-order normalisation condition. In order to diagonalise
$\rho_{AB}$ and $\rho_{AD}$ we use the bases 
$\{ |u_{\pm }\ra ,|v_{\pm }\ra \}$ and
$\{ |w_{\pm }\ra ,|x_{\pm }\ra  \}$, respectively, where 
\begin{align*}
&|u_{\pm }\ra =( \sqrt{2}\mp 1 )|10\ra\pm |00\ra, 
&&|v_{\pm }\ra = (\sqrt{2}\pm 1)|11\ra \mp |01\ra, \\
&|w_{\pm }\ra =(\sqrt{2}\mp 1 )|10\ra \mp |00\ra, 
&&|x_{\pm }\ra = ( \sqrt{2}\pm 1 )|11\ra\pm |01\ra.
\end{align*}
Therefore we can now write
\begin{align*}
\rho_{AB}=\rho_{AD} &=\frac{1}{8} 
\begin{bmatrix}
2+\sqrt{2} & & & \\
& 2-\sqrt{2} && \\
&& 2+\sqrt{2} & \\
&&&2-\sqrt{2} 
\end{bmatrix}\\
&=(2-\sqrt{2})\frac{\mathbb{I}}{8}+\frac{\sqrt{2}}{4}\begin{bmatrix}
1 & & & \\
& 0& & \\
&& 1 & \\
&&&0 
\end{bmatrix}.
\end{align*}
Hence, by (\ref{firstorder}) and the normalisation condition, 
\begin{align*}
\delta E_{2} &\propto \tr [\delta \rho_{AB}
(|u_{+} \ra \la u_{+}|+|v_{+} \ra \la v_{+}|)+\delta \rho_{AD}(|w_{+} \ra \la w_{+}|+|x_{+} \ra
\la x_{+}|)]\\
	&=\la u_{+}|\delta \rho_{AB} |u_{+}\ra +\la v_{+}|\delta \rho_{AB}
|v_{+}\ra +\la w_{+}|\delta \rho_{AD} |w_{+}\ra + \la x_{+}|\delta \rho_{AD} |x_{+}\ra.
\end{align*}
Consider a variation of the form 
$$|\delta \psi \ra = \alpha |0011\ra +\beta |1011\ra.$$
Then the normalisation condition requires that $\re 
\{\alpha +\beta \}=0$. We find
$$\delta \rho_{AB}=\frac{1}{\sqrt{2}} (|00\ra+|10\ra )
(\overline{\alpha} \la00| +\overline{\beta}
\la10|)
+ \frac{1}{\sqrt{2}}(\alpha|00\ra + \beta|10\ra)(\la 00| + \la 10|),$$ 
and hence
\begin{align*}
\la u_{+}|\delta \rho_{AB} | u_{+} \ra 	&=
2\,{\rm Re}\,\left[\alpha +\beta  (\sqrt{2}-1)\right],\\
\la v_{+}|\delta \rho_{AB} | v_{+} \ra 		&=0.
\end{align*} 
Similarly, 
\begin{align*}
\la w_{+}|\delta \rho_{AD} | w_{+} \ra 	&=0, \\
\la x_{+}|\delta \rho_{AD} | x_{+} \ra 	&=2\,{\rm Re}\,\left[
\alpha(\sqrt{2}+1) +\beta  (2\sqrt{2}+3)\right].
\end{align*} 
Therefore, for the variations considered here we have 
$$\delta E_{2}\propto {\rm Re}\,\left[
\alpha (2+\sqrt{2})+\beta (2+3\sqrt{2})\right].$$
Hence, by putting $\alpha =-\beta =\varepsilon $ for a small 
$\varepsilon \in \mathbb{R}$ --- note that the normalisation condition
$\re (\alpha+\beta)=0$ is satisfied ---
we have $\delta E_2 \neq 0$. 
Therefore the state $|\psi_{4}\ra$ cannot give a local maximum of 
$E_{2}$.

			\section{The Second-order variations}

We now return to the state $|M_{4}\ra$ and 
show that it gives a local maximum of $E_2$,
that is, $\delta^{2} E_{AB}+\delta^{2} E_{AC}+\delta^{2} E_{AD}<0.$ 
Let us write $\delta^{2}\rho_{XY}=\kappa_{XY}+\sigma_{XY},$ where
\begin{align*}
\sigma_{XY} &=\tr_{WZ}(|\delta\psi\ra\la\delta\psi|),\\
\kappa_{XY} &=\tr_{WZ}(|\delta^{2}\psi\ra\la\psi|+|\psi\ra\la 
\delta^{2} \psi|).
\end{align*}
Then,
\begin{equation}
- \log 2\sum_{Y} \delta^{2} E_{AY}=
\sum_{Y}\tr(\kappa_{AY}\log\rho_{AY}+\sigma_{AY}
\log\rho_{AY} +\frac{1}{2}\delta\rho_{AY}\delta\log\rho_{AY}),
\label{sumY}
\end{equation}
where $Y = B,C$ and $D$. 
Our task now is to show that the right-hand side of this equation 
is positive definite for all nontrivial 
variations of the state $|\psi\ra = |M_4\ra$ satisfying the
normalisation conditions (\ref{norm 1}) and (\ref{norm 2}). 
We will deal with each term in (\ref{sumY}) separately.

	\subsection{The first two terms in the expansion}

We note that $\log\rho_{AY}=\log3\cdot|\Phi_{-}\rangle\langle 
\Phi_{-}|-\log6\cdot \mathbb{I}$, where
$
|\Phi_{-} \ra = \frac{1}{\sqrt{2}}(|10\ra - |01\ra)
$,
for all $Y$, and that  
\begin{align*}
\tr(\kappa_{AB}|\Phi_{-}\rangle\langle\Phi_{-}|)
&=\textrm{Re}(\langle\delta^{2}
\psi|\psi\rangle-\langle\delta^{2}\psi|\overline{\psi}\rangle),\\
\tr(\kappa_{AC}|\Phi_{-}\rangle\langle \Phi_{-}|)
&=\textrm{Re}(\langle\delta^{2}\psi|\psi\rangle
-\overline{\omega}\langle\delta^{2}\psi|\overline{\psi}\rangle), \\
\tr(\kappa_{AD}|\Phi_{-}\rangle\langle\Phi_{-}|)	
&=\textrm{Re}(\langle\delta^{2}
\psi|\psi\rangle-\omega\langle\delta^{2}\psi|\overline{\psi}\rangle),
\end{align*}
where $|\overline{\psi}\ra$ is the complex conjugate of 
$|\psi\ra = |M_4\ra$ in the computation basis, 
as was shown in~\cite{HS00} in the context of first-order variation.
Hence, using the normalisation condition (\ref{norm 2}), we find the
first term of (\ref{sumY}) as $$
\sum_{Y}\tr(\kappa_{AY}\log\rho_{AY})
=3\log 2\sqrt{3}\cdot\langle\delta\psi|\delta\psi\rangle.$$

We now consider the second term in (\ref{sumY}).  We have
\begin{align*}
\tr(\sigma_{XY}\log\rho_{XY})	&=\tr(\sigma_{XY}\log3\cdot
|\Phi_{-}\rangle\langle\Phi_{-}|-\sigma_{XY}\log6\cdot \mathbb{I})\\
&=\log 3\cdot\tr[(\tr_{WZ}|\delta\psi\rangle\langle\delta\psi|)
|\Phi_{-}\rangle\langle\Phi_{-}|]-\log6\cdot
\langle\delta\psi|\delta\psi\rangle\\
&=\log{3} \sum ^{1}_{i,j=0}|(\la\Phi _{-}|_{XY}\la 
ij|_{WZ})|\delta\psi\ra|^{2}-\log6\cdot
\langle\delta\psi|\delta\psi\rangle.
\end{align*}
Hence,
$$ \sum _{Y} \tr(\sigma_{AY}\log\rho_{AY})
=-3\log 6\cdot
\langle\delta\psi|\delta\psi\rangle + F_{AB}+F_{AC}+F_{AD},$$
where 
$$F_{XY}=\log{3} 
\sum ^{1}_{i,j=0}|(\la\Phi _{-}|_{XY}\la ij|_{WZ})
|\delta\psi\ra|^{2}.$$
Thus
\begin{eqnarray}
\sum_{Y}\tr\left[(\kappa_{AY}+\sigma_{AY})\log\rho_{AY}\right]
& = & -3 \log {\sqrt{3}} \cdot \la \delta\psi | \delta \psi \ra   
+ F_{AB}+F_{AC}+F_{AD}\nonumber \\
& \geq & - \frac{3}{2}\log 3\cdot \la\delta \psi|\delta
\psi\ra\,,\nonumber 
\end{eqnarray}
because
$F_{AB}+F_{AC}+F_{AD}\geq 0$.  This motivates us to define
\begin{equation}
P \equiv \sum_{Y} \tr[\delta\rho_{AY}\delta \log \rho_{AY}]
- 3\log 3\cdot\la \delta\psi|\delta \psi\ra\,. \label{thisisit}
\end{equation}
Then, if $P>0$ for all nontrivial variations, $\delta^2 E_2$ is
negative definite and the state $|M_4\ra$ gives a local maximum
of $E_2$.  We will show this fact with a certain convenient
parametrisation of variations.

\subsection{The third term in the expansion} 
The following lemma will be useful in analysing the variation
$\delta \log \rho_{XY}$.
\begin{lem}\label{logA}
Provided that the eigenvalues of $A$ are positive and less than $1$,
we have
$$\delta \log A=\int^{1}_{0} [\mathbb{I}-t(\mathbb{I}-A)]^{-1}
\delta A[\mathbb{I}-t(\mathbb{I}-A)]^{-1}dt.$$
\begin{proof}
We expand $\log A$ as
$$
\log A  = \log (\mathbb{I}-(\mathbb{I}-A)) 
=-\sum_{n=1}^{\infty}\frac{(\mathbb{I}-A)^{n}}{n}.
$$
Then
\begin{align*}
\delta \log A 	
&=\sum_{n=1}^{\infty}\sum_{m=0}^{n-1} \frac{(\mathbb{I}-A)^{m}
\delta A(\mathbb{I}-A)^{n-m-1}}{n} \\
&=\sum_{n'=0}^{\infty}\sum_{m=0}^{\infty} \frac{(\mathbb{I}-A)^{m}
\delta A(\mathbb{I}-A)^{n'}}{n'+m+1},
\end{align*}
where we have let $n'=n-m-1$. Noting the elementary integral,
$$\int^{1}_{0}t^{n'+m}dt =\frac{1}{n'+m+1},$$
we find
\begin{align*}
\delta \log A &=\int^{1}_{0}
\sum_{n'=0}^{\infty}\sum_{m=0}^{\infty}
[t(\mathbb{I}-A)]^{m}\delta A[t(\mathbb{I}-A)]^{n'} dt \\
    &=\int^{1}_{0}
\sum_{m=0}^{\infty} [t(\mathbb{I}-A)]^{m}\delta
A\sum_{n'=0}^{\infty}[t(\mathbb{I}-A)]^{n'} dt \\
    &=\int^{1}_{0}
[\mathbb{I}-t(\mathbb{I}-A)]^{-1}
\delta A[\mathbb{I}-t(\mathbb{I}-A)]^{-1} dt,
\end{align*}
as required.
\end{proof}
\end{lem}
We use this lemma with $A=\rho$, 
where $\rho$ is a density matrix. Since $\rho$ is
Hermitian, we can choose a basis in which
$\rho=\textrm{diag}\{\lambda _{1}, \ldots,\lambda _{n}\}$. 
We can apply this lemma if $0<\lambda _{i}<1$ for all $i$.
Note that the state $|M_4\rangle$ has this property.
If $\lambda_i\neq \lambda_j$, we have
\begin{align*}
(\delta \log \rho)_{ij}	&=\int^{1}_{0} 
\frac{\delta_{ik}}{1-t(1-\lambda_{i})} \delta \rho_{kl} 
\frac{\delta_{lj}}{1-t(1-\lambda_{j})} dt \\
&=\int^{1}_{0}\left (\frac{X}{1-t(1-\lambda_{i})}+\frac{Y}{1-t(1-\lambda_{j})}\right)
\delta \rho_{ij}dt ,
\end{align*}
where
$X=\frac{1-\lambda_{i}}{\lambda_{j}-\lambda_{i}}$ and
$Y=\frac{1-\lambda_{j}}{\lambda_{i}-\lambda_{j}}$. 
Hence,
for $\lambda_{i}\neq\lambda_{j}$,
\begin{equation}
(\delta \log \rho)_{ij}	
= 
\frac{1}{\lambda_{j}-\lambda_{i}}\log\left (\frac{\lambda_{j}}{\lambda_{i}}\right)
\delta \rho_{ij}. \label{differ}
\end{equation}
If $\lambda_{i}=\lambda_{j}$, then
\begin{eqnarray}
(\delta \log \rho)_{ij}	
&=&\int^{1}_{0} \frac{\delta \rho_{ij}}{(1-t(1-\lambda_{i}))^{2}} dt
\nonumber \\
&=& \frac{\delta \rho_{ij}}{\lambda_{i}}. \label{same}
\end{eqnarray}
(This formula 
can also be obtained by letting $\lambda_j \to \lambda_i$ in 
(\ref{differ}).) 
We now apply these formulas to the variation $\delta\log \rho_{XY}$.

To ease the notation we let $\delta\rho_{AB}=(a_{ij}^{(1)})$,
$\delta\rho_{AC}=(a_{ij}^{(2)})$ and
$\delta\rho_{AD}=(a_{ij}^{(3)})$.  We will use the basis
$S=\{|O\rangle,|I\rangle,|+\rangle,|-\rangle
\}=\{|00\rangle,|11\rangle,|\Phi_{+}\rangle,|\Phi_{-}\rangle \}$,
where
$|\Phi_{\pm}\rangle = 2^{-1/2}(|10\rangle - |01\rangle)$.
Thus, for example, $a^{(1)}_{OI}=\langle O|\delta\rho_{AB}|I\rangle$. 
Since $ \lambda _{1}=\lambda _{2}=\lambda _{3}=1/6$ and $\lambda _{4}=1/2$
for each $\rho_{AY}$, we have, by applying (\ref{differ}) and
(\ref{same}),
\begin{align*}
	\tr(\delta\rho_{AB}\delta\log\rho_{AB})
&=\sum_{ij}K_{ij}|a^{(1)}_{ij}|^{2}, \\
    \tr(\delta\rho_{AC}\delta\log\rho_{AC})
&=\sum_{ij}K_{ij}|a^{(2)}_{ij}|^{2},\\
    \tr(\delta\rho_{AD}\delta\log\rho_{AD})&=\sum_{ij}K_{ij}|
a^{(3)}_{ij}|^{2},
\end{align*}
where $$K=\left(%
\begin{array}{cccc}
  6 & 6 & 6 & 3\log3 \\
  6 & 6 & 6 & 3\log3 \\
  6 & 6 & 6 & 3\log3 \\
  3\log3 & 3\log3 & 3\log3 & 2 \\
\end{array}%
\right).$$ 
				
In order to proceed further, we need to explicitly parametrise 
the variations $|\delta\Psi\rangle$.  
Thus, we write
\begin{eqnarray}
|\delta\Psi\rangle & = & 
\epsilon_{0000}|0000\rangle+ \epsilon_{1111}|1111\rangle \nonumber \\
&& + \epsilon_{0011}|0011\rangle + \epsilon_{1100}|1100\rangle
+ \omega(\epsilon_{1010}|1010\rangle + \epsilon_{0101}|0101\rangle)
\nonumber \\
&& + \omega^2(\epsilon_{1001}|1001\rangle + \epsilon_{0110}
|0110\rangle)\nonumber \\
&& + \epsilon_{0111}|0111\rangle
+ \epsilon_{1011}|1011\rangle + \epsilon_{1101}|1101\rangle
+ \epsilon_{1110}|1110\rangle\nonumber\\
&& + z(|1000\rangle + |0100\rangle + |0010\rangle + |0001\rangle) \,.
\nonumber
\end{eqnarray}
Note that we have included only 
one term with ``three $0$'s" out of four possible
terms. This is because all other terms can be eliminated by local
unitary transformations to first order.
We derive additional constraints on the variations by noting 
that if the effect of the variation
is to change the relative phase in
any one of the qubits, then our new state 
$|\psi^{\prime}\ra$ is locally
equivalent to $|M_{4}\ra$. 
Let us write 
\begin{align*}
\epsilon_{0011}-\epsilon_{1100} &=x_{1}+iy_{1},\,\,\,\,\,
\epsilon _{1100}+\epsilon _{0011}   =X_{1}+iY_{1},\\
\epsilon_{0101}-\epsilon_{1010} &=x_{2}+iy_{2},\,\,\,\,\,
\epsilon _{1010}+\epsilon _{0101}   =X_{2}+iY_{2},\\
\epsilon_{0110}-\epsilon_{1001} &=x_{3}+iy_{3},\,\,\,\,\,
\epsilon _{1001}+\epsilon _{0110}   =X_{3}+iY_{3}.
\end{align*} 
The first-order 
variation in the relative phase within the first qubit results in
the change in $y_1+y_2+y_3$.  Similarly, the phase variations in the
second and third qubits change the values of $-y_1+y_2+y_3$ and
$y_1-y_2+y_3$, respectively.  Hence, for any variation, we can always
find an equivalent variation satisfying
\begin{equation}\label{phase}
y_{1}=y_{2}=y_{3}=0, \end{equation}
by adjusting these phases.
In the same way
we find that an overall change of phase,
$|M_{4}\ra\rightarrow e^{i\theta }|M_{4}\ra$,
can be used to impose the condition
\begin{equation}\label{phase2}
Y_{1}+Y_{2}+Y_{3}=0.
\end{equation}
Thus, we have $21$ real parameters (after imposing the normalisation
condition) in our space of variations.  Since the dimensionality 
of the space of locally
inequivalent states is $18$ (see, e.g.~\cite{NP98}) in a neighbourhood
of a generic state, three
dimensions are redundant.  This discrepancy is due to the fact
that the state $|M_4\rangle$ remains 
unchanged if all qubits are transformed
by the same $SU(2)$ matrix: thus, the dimensionality of 
the orbit of the local unitary
transformations of the state $|M_4\rangle$ 
is $10$, which is smaller than the dimensionality of this orbit for
a generic state by $3$.  We have eliminated all variations 
that reduce to
infinitesimal local unitary transformations, but 
the set of physically equivalent variations is generically three
dimensional.  It would be possible to 
eliminate this redundancy
by using canonical forms~\cite{CHS,AAJT01} though we 
have chosen not to do so. 

With the condition (\ref{phase}) imposed, the quantity $P$ defined by
(\ref{thisisit}) can be written as
$$
P = P_1 + P_2 + P_3 + P_4\,,
$$
where
\begin{eqnarray}
P_1 & = & 12\sum_{\alpha=1}^3 |a_{OI}^{(\alpha)}|^2
- 3\log 3(|\epsilon_{0000}|^2 + |\epsilon_{1111}|^2)\,,\label{P1}\\
P_2 & = & \sum_{\alpha=1}^3
\left[ 12(|a_{I+}^{(\alpha)}|^2+|a_{O+}^{(\alpha)}|^2)
+ 6\log 3(|a_{I-}^{(\alpha)}|^2 + |a_{O-}^{(\alpha)}|^2)\right]\nonumber
\\
&&
- 3\log 3(|\epsilon_{0111}|^2 + |\epsilon_{1011}|^2 +
|\epsilon_{1101}|^2
+ |\epsilon_{1110}|^2 + 4|z|^2)\,,\label{P2}\\
P_3 & = & 6\log 3\sum_{\alpha=1}^3|a_{+-}^{(\alpha)}|^2
- \frac{3}{2}\log 3 (x_1^2+x_2^2+x_3^2)\,,\label{P3}\\
P_4 & = & \sum_{\alpha=1}^3
\left[ 6(|a_{OO}^{(\alpha)}|^2 + |a_{II}^{(\alpha)}|^2
+ |a_{++}^{(\alpha)}|^2) + 2|a_{--}^{(\alpha)}|^2\right]
- \frac{3}{2}\log 3\sum_{\alpha=1}^3 (X_\alpha^2 +
Y_\alpha^2)\,.\nonumber \\
\label{P4}
\end{eqnarray} 
We will show that (i) $P_1> 0$ if either $\epsilon_{0000}$ or
$\epsilon_{1111}$ is nonzero, (ii) $P_2 > 0$ if any of
$\epsilon_{0111}$, $\epsilon_{1011}$, $\epsilon_{1101}$,
$\epsilon_{1110}$ or $z$ is nonzero, 
(iii) $P_3 > 0$ if any of $x_\alpha$'s is
nonzero and (iv) $P_4>0$ if any or $X_\alpha$'s or $Y_\alpha$'s is
nonzero.  This will imply that $P$ is positive definite.

\subsubsection{Positivity of $P_1$}
The only terms relevant here are $a^{(\alpha)}_{OI}$'s.
These are given by
\begin{align*}
\sqrt{6}a^{(1)}_{OI}=\langle O|\delta\rho_{AB}|I\rangle
&=\langle\delta\psi|II\rangle+\langle OO|\delta\psi\rangle
=\overline{\epsilon_{1111}}+\epsilon_{0000},\\
\sqrt{6}a^{(2)}_{OI}=\langle O|\delta\rho_{AC}|I\rangle
&=\omega\langle\delta\psi|II\rangle+\overline{\omega}\langle OO|\delta\psi\rangle
=\omega\overline{\epsilon_{1111}}+\overline{\omega}\epsilon_{0000},\\
\sqrt{6}a^{(3)}_{OI}=\langle O|\delta\rho_{AD}|I\rangle
&=\overline{\omega}\langle\delta\psi|II\rangle+\omega\langle
OO|\delta\psi\rangle
=\overline{\omega}\overline{\epsilon_{1111}}+\omega\epsilon_{0000}.
\end{align*}
Thus we have
$$
\sum_{\alpha=1}^3 |a_{OI}^{(\alpha)}|^2 =
\frac{1}{2}(|\epsilon_{0000}|^2 + |\epsilon_{1111}|^2)\,.
$$
Hence by (\ref{P1})
$$
P_1 = (6-3\log 3)(|\epsilon_{0000}|^2 + |\epsilon_{1111}|^2)\,,
$$
which is positive if either $\epsilon_{0000}$ or $\epsilon_{1111}$ is
nonzero.

\subsubsection{Positivity of $P_2$}
We find the relevant $a_{ij}^{(1)}$'s as
\begin{eqnarray}
\sqrt{6}\,a_{I+}^{(1)} & = & -\frac{1}{\sqrt{2}}(\epsilon_{1101}+
\epsilon_{1110}) + \sqrt{2}\,\overline{z},\nonumber \\
\sqrt{6}\,a_{I-}^{(1)} & = & -\sqrt{\frac{3}{2}}i\,(\epsilon_{1110}-
\epsilon_{1101}),\nonumber \\
\sqrt{6}\,a_{O+}^{(1)} & = & \frac{1}{\sqrt{2}}(\overline{\epsilon_{0111}}
+\overline{\epsilon_{1011}}) - \sqrt{2}\,z,\nonumber \\
\sqrt{6}\,a_{O-}^{(1)} & = & \frac{1}{\sqrt{2}}(\overline{\epsilon_{1011}}
- \overline{\epsilon_{0111}})\,.\nonumber
\end{eqnarray}
Hence,
\begin{eqnarray}
12|a_{I+}^{(1)}|^2 + 6\log 3|a_{I-}^{(1)}|^2
& = & |\epsilon_{1101}+\epsilon_{1110}-2\overline{z}|^2
+ \frac{3}{2}\log 3|\epsilon_{1101}-\epsilon_{1110}|^2\,,\nonumber \\
12|a_{O+}^{(1)}|^2 + 6\log 3|a_{O-}^{(1)}|^2
& = & |\epsilon_{1011}+\epsilon_{0111}-2\overline{z}|^2
+ \frac{1}{2}\log 3|\epsilon_{1011} - \epsilon_{0111}|^2\,.
\nonumber
\end{eqnarray}
The corresponding quantities involving $a^{(2)}_{ij}$'s and
$a^{(3)}_{ij}$ can be obtained similary as follows:
\begin{eqnarray}
12|a_{I+}^{(2)}|^2 + 6\log 3|a_{I-}^{(2)}|^2
& = & |\epsilon_{1011}+\epsilon_{1110}-2\omega^2\overline{z}|^2
+ \frac{3}{2}\log 3|\epsilon_{1011}-\epsilon_{1110}|^2\,,\nonumber \\
12|a_{O+}^{(2)}|^2 + 6\log 3|a_{O-}^{(2)}|^2
& = & |\epsilon_{1101}+\epsilon_{0111}-2\omega^2\overline{z}|^2
+ \frac{1}{2}\log 3|\epsilon_{1101} - \epsilon_{0111}|^2\,,
\nonumber \\
12|a_{I+}^{(3)}|^2 + 6\log 3|a_{I-}^{(3)}|^2
& = & |\epsilon_{1011}+\epsilon_{1101}-2\omega\overline{z}|^2
+ \frac{3}{2}\log 3|\epsilon_{1011}-\epsilon_{1101}|^2\,,\nonumber \\
12|a_{O+}^{(3)}|^2 + 6\log 3|a_{O-}^{(3)}|^2
& = & |\epsilon_{1110}+\epsilon_{0111}-2\omega\overline{z}|^2
+ \frac{1}{2}\log 3|\epsilon_{1110} - \epsilon_{0111}|^2\,.
\nonumber
\end{eqnarray}
{}From these equations we find
\begin{eqnarray}
P_2 & = &
\left( 1- \frac{1}{2}\log 3\right)(|\epsilon_{0111}+\epsilon_{1011}|^2 
+ |\epsilon_{0111}+\epsilon_{1101}|^2 + 
|\epsilon_{0111} + \epsilon_{1110}|^2)\nonumber \\
&& 
+\left( 1- \frac{1}{2}\log 3\right)(|\epsilon_{1011}+\epsilon_{1101}|^2 
+ |\epsilon_{1011}+\epsilon_{1110}|^2 + 
|\epsilon_{1101} + \epsilon_{1110}|^2)\nonumber \\
&& + \log 3(|\epsilon_{1011} - \epsilon_{1101}|^2
+ |\epsilon_{1011}-\epsilon_{1110}|^2
+ |\epsilon_{1101}-\epsilon_{1110}|^2)\nonumber \\
&& + 24\left(1-\frac{1}{2}\log 3\right)|z|^2\,.\nonumber
\end{eqnarray}
It is clear that the right-hand side
is positive unless $\epsilon_{0111}$,
$\epsilon_{1011}$, $\epsilon_{1101}$, $\epsilon_{1110}$ and $z$ all 
vanish.\footnote{The expression for $P_2$ is not symmetric under
permutations of four qubits involving the first qubit.  However,
$\delta^2E_2$ itself is symmetric under such permutations thanks to the
contribution from $F_{AB}+F_{AC}+F_{AD}$ (which we have discarded
because it is positive definite).
This must be the case because the average two-partite von Neumann
entanglement entropy $E_2$ has this symmetry.}

\subsubsection{Positivity of $P_3$}
We have
$$
2\sqrt{6}\,a_{+-}^{(1)} = -2(x_2+x_3) - \sqrt{3}i(x_2-x_3),
$$
and $a_{+-}^{(2)}$ and $a_{+-}^{(3)}$ are obtained from this by cyclic permutations
$2\to 3\to 1$ and $3\to 2 \to 1$, respectively.  Hence,
$$
|a_{+-}^{(1)}|^2 + |a_{+-}^{(2)}|^2 + |a_{+-}^{(3)}|^2
= \frac{1}{12}(7x_1^2+7x_2^2+7x_3^2 + x_2x_3 + x_3x_1 + x_1x_2)\,.
$$
Thus, $P_3$ given by (\ref{P3}) is
$$
P_3 = \frac{\log 3}{2}(4x_1^2+4x_2^2+4x_3^2 + x_2x_3+x_3x_1 + x_1x_2)\,,
$$
which is positive if $x_1$, $x_2$ or $x_3$ is nonzero.

\subsubsection{Positivity of $P_4$}
We have
\begin{align*}
\sqrt{6}a_{OO}^{(1)}&
=\epsilon_{0011}+\overline{\epsilon_{0011}} =X_{1}+x_{1},\\
\sqrt{6}a_{II}^{(1)}&
    =\epsilon_{1100}+\overline{\epsilon_{1100}}
    =X_{1}-x_{1}.
\end{align*}
Hence
$$
6|a_{OO}^{(1)}|^{2}+6|a_{II}^{(1)}|^{2}=
2(X_{1}^{2}+x_{1}^{2}).
$$
Similarly,
\begin{align*}
6|a_{OO}^{(2)}|^{2}+6|a_{II}^{(2)}|^{2}&=2(X_{2}^{2}+x_{2}^{2}),\\
6|a_{OO}^{(3)}|^{2}+6|a_{II}^{(3)}|^{2}&=2(X_{3}^{2}+x_{3}^{2}).
\end{align*}
Thus,
\begin{equation}
6\sum_{\alpha=1}^3(|a_{OO}^{(\alpha)}|^2 + |a_{II}^{(\alpha)}|^2)
\geq 2\sum_{\alpha=1}^3 X_\alpha^2. \label{1st}
\end{equation}
The remaining `diagonal terms' are
\begin{align*}
-\sqrt{6}a_{++}^{(1)}
 &=\textrm{Re}
\{\omega(\epsilon_{1010}+\epsilon_{0101})+\overline{\omega}
(\epsilon_{0110}+\epsilon_{1001})\},\\
-\sqrt{2}a_{--}^{(1)}
    &=\textrm{Im} \{\omega(\epsilon_{1010}+\epsilon_{0101})
-\overline{\omega}(\epsilon_{0110}+\epsilon_{1001})\}.
\end{align*}
The coefficients $a_{++}^{(2)}$ and $a_{--}^{(2)}$ ($a_{++}^{(3)}$ and
$a_{--}^{(3)}$) are obtained from the expressions for 
$a_{++}^{(1)}$ and $a_{--}^{(1)}$ by interchanging the second and third
(fourth) qubits.
Remembering the definitions
$\epsilon_{1010} + \epsilon_{0101} = X_2 + iY_2$ and
$\epsilon_{1001} + \epsilon_{0110} = X_3 + iY_3$, we obtain
$$
6|a_{++}^{(1)}|^2 + 2|a_{--}^{(1)}|^2
= X_2^2 + Y_2^2 + X_3^2 + Y_3^2 + 2(X_2X_3-Y_2Y_3).
$$
We find similarly
\begin{eqnarray}
6|a_{++}^{(2)}|^2 + 2|a_{--}^{(2)}|^2
& = & X_3^2 + Y_3^2 + X_1^2 + Y_1^2 + 2(X_3X_1-Y_3Y_1),\nonumber\\
6|a_{++}^{(3)}|^2 + 2|a_{--}^{(3)}|^2
& = & X_1^2 + Y_1^2 + X_2^2 + Y_2^2 + 2(X_1X_2-Y_1Y_2).\nonumber
\end{eqnarray}
By combining these formulas and (\ref{1st}) with the definition
(\ref{P4}) of $P_4$ we have
\begin{eqnarray}
P_4 & \geq &\left(4-\frac{3}{2}\log 3\right)(X_1^2+X_2^2+X_3^2)
+ 2(X_2X_3+X_3X_1+X_1X_2) \nonumber \\
&& \ \ \ + \left( 2 -\frac{3}{2}\log 3\right)(Y_1^2+Y_2^2+Y_3^2)
- 2(Y_2Y_3 + Y_3Y_1 + Y_1Y_2).  \nonumber
\end{eqnarray}
Finally, the use of the condition $Y_1+Y_2+Y_3=0$ leads to
\begin{eqnarray}
P_4 & \geq  & \left[ (X_2+X_3)^2 + (X_3+X_1)^2 +
(X_1+X_2)^2\right]\nonumber \\
&& + \left(2-\frac{3}{2}\log 3\right)(X_1^2+X_2^2+X_3^2) + 
3\left(2 - \log 3\right)
(Y_1^2 + Y_1Y_2 + Y_2^2).\nonumber
\end{eqnarray}
Thus, $P_4> 0$ unless $X_\alpha$'s and $Y_\alpha$'s 
all vanish. This completes the
proof that $\delta^{2}E_{2}$ at
$|M_{4}\ra $ is negative definite.\footnote{It is possible to prove
$\delta^2E_2 < 0$ without the condition $Y_1+Y_2+Y_3 = 0$.  In that
case, one needs to evaluate the positive contribution $F_{AB} +
F_{AC} + F_{AD}$ to $\sum_{Y}\tr[\sigma_{AY}\log\rho_{AY}]$.}
Hence, the state $|M_{4} \ra $ 
indeed gives a local maximum of the average two-partite von Neumann 
entanglement entropy.

\end{document}